\documentclass[12pt]{article}
\usepackage[pctex32]{graphics}
\begin{document}
~
\begin{center} {\Large \bf  Giant Magnons under NS-NS and Melvin Fields}
                                                  
\vspace{1cm}

                      Wung-Hong Huang\\
                       Department of Physics\\
                       National Cheng Kung University\\
                       Tainan, Taiwan\\

\end{center}
\vspace{1cm}
\begin{center} {\large \bf  Abstract} \end{center}
The giant magnon is a rotating spiky string configuration which has the same dispersion relation between the energy and angular momentum as that of a spin magnon.  In this paper we investigate the effects of the NS-NS and Melvin fields on the giant magnon. We first analyze the energy and angular momenta of the two-spin spiky D-string moving on the $AdS_3\times S^1$ with the NS-NS field.   Due to the infinite boundary of the AdS spacetime the D-string solution will extend to infinity and it appears the divergences.  After adding the counter terms we obtain the dispersion relation of the corresponding giant magnon.  The result  shows that there will appear a prefactor before the angular momentum, in addition to some corrections in the sine function.  We also see that the spiky profile of a rotating D-string plays an important role in mapping it to a spin magnon.  We next investigate the energy and angular momentum of the one-spin spiky fundamental string moving on the $R \times S^2$ with the electric or magnetic Melvin field.  The dispersion relation of the corresponding deformed giant magnon is also obtained.   We discuss some properties of the correction terms and their relations to the spin chain with deformations.
\vspace{3cm}
\begin{flushleft}
E-mail:  whhwung@mail.ncku.edu.tw\\
\end{flushleft}
%%%%%%%%%%%%%%%%%%%%%%%%%%%%%%%
\newpage
\section {Introduction}
It is known that the AdS/CFT correspondence plays important role in studying the gauge theories at strong coupling [1-3].   The correspondence 
is the equivalence between the spectrum of free string theory on $AdS_5\times S^ 5$ and the spectrum of  anomalous dimensions of gauge invariant operators in the N = 4 supersymmetric Yang-Mills (SYM) theory.  The most studied cases were in the long-wave approximation corresponding to classical rotating and pulsating strings [4-6].  Another important cases are the low lying spin chain states corresponding to the magnon excitations.  In recent,  Maldacena and Hofman [7] had identified the elementary magnon with a rotating spiky string configuration moving on an $R \times S^2$, which they called the `giant magnon'.   The dispersion relation between energy $E$ and angular momentum $J$ for the one spin giant magnon  is  
$$E - J = {\sqrt\lambda\over \pi}\,|\sin{p\over2}\,|\,,\eqno{(1.1)}$$
where $p$ is the magnon momentum, which on the string side is interpreted as a geometrical angle of the string.   Using the correspondence between the giant magnon and the sine-Gordon soliton [8] they had calculated the scattering phase of two magnons and shown that it matches the large  $\lambda$ limit of the conjecture of [9].   

In the subsequent works [10], Dorey et al constructed the classical string solution which corresponds to the bound state of the giant magnon.   The two-charge giant magnon had also been found in [10] which exploits the correspondence between classical string and the complex sine-Gordon model.   More properties on giant magnons have been investigated, in which the finite $J$ effects [11], quantum corrections [12], multispin properties [13,14], magnon from M-theory [15] and giant magnon moving on the $\beta$ deformed $AdS_5 \times S^5$ [16,14] are discussed.

The aim of this paper is to investigate the spiky D-string moving on the $AdS_3\times S^1$ with NS-NS antisymmetric B field background which was first used by Bachas and Petropoulos to analyze the Anti-de-Sitter D-brane [17-19]. The background is the near-horizon geometry of a black string which is constructed out of NS five-branes and fundamental string, in contrast to the  $\beta$-deformation background [20] which related to ${\cal N} =1 $ $\beta$ deformed SYM.   We will also investigate the spiky fundamental string moving on the electric/magnetic Melvin fields background which was constructed by us and had been used to analyze the properties of classical string solution and giant graviton therein [21,22]. As the magnetic fields couple differently to particles of different spins they naturally break supersymmetry.    Therefore the correspondence between the magnon and string will be in the content of less supersymmetry, which is of interesting from  the point of view of phenomena. 

As the giant magnon discussed in [7] is a spiky string soliton solution we will first in section II follow the method of  [23] to investigate the spiky D-string in the flat spacetime with external flux field.  It is used to see how the field will modify the Regge trajectory.   In section III we follow the method of  [12] to evaluate the energy and angular momenta of the two-spin spiky D-string moving on the $AdS_3\times S^1$ with NS-NS B field [17].   Due to the infinite boundary of the AdS spacetime the solution will extend to infinity and it will appear the divergences.  We adopt the prescription of [12] to add the counter term to obtain a finite result of dispersion relation of the corresponding giant magnon.   The result is used to find how the background field will correct the relation.  We also see that a spiky profile of a rotating D-string is crucial to map it to a magnon.  In section IV  we follow the method of  [11] to evaluate the energy and angular momentum of  the one-spin spiky fundamental string moving on the electric or magnetic Melvin-field deformed $R \times S^2$.  We also find the dispersion relations of the corresponding Melvin-field deformed giant magnons.   We mention the possible relation of the correction terms in the dispersion relation to the deformations of the spin chain [24-26].  The last section is devoted to the discussion. 

  Note that as the analyses in sections II and III will follow [12] in which the Nambu-Goto action is adopted, therefore under the NS-NS B field we will adopt the Langrangian $\sim \sqrt{det(g_{ab}+B_{ab})}$ which will describe a D-string.   On the other hand, in the section IV, as we follow the method in [11] in which the Polyakov action is adopted. Therefore under the NS-NS B field we simply add a term $\sim B_{\mu\nu}\epsilon^{ab}\partial_a X^{\mu} \partial_b X^{\nu}$ to the Langrangian and it will describe a fundamental string.  

%%%%%%%%%%%%%%%%%%%%%
\section {Spiky D-String under Flux Field}
In this section we will follow the method of  [23] to investigate the spiky D-string in the flat spacetime with flux field.  We consider the D-string moving on the flat x-y plan with a constant NS-NS or EM flux field, $B_{xy}=B$.   In terms of the polar coordinate the line element and B field are described as
$$ds^2= - dt^2 +d\rho^2 + \rho^2 d\theta^2,~~~~~~~~~~~B_{\rho\theta} = B\rho.  \hspace{3cm}\eqno{(2.1)}$$
To consider a rotating D-string we adopt the world-sheet coordinate [23] such that 
$$t = \tau, ~~~~~~\theta = \omega\tau +\sigma, ~~~~~\rho=\rho(\sigma).\eqno{(2.2)}$$
The Nambu-Goto Lagrangian of the rotating D-string is described by 
$${\cal L} = {\sqrt\lambda\over 2\pi}\int d\sigma \sqrt{\left(1-(1+B^2)\omega^2\rho^2\right)\rho'^2+\rho^2}.\eqno{(2.3)}$$
The equation of motion of the D-string is described by the equation 
$$\rho' = {\rho\rho_1\over\rho_0}{\sqrt{\rho^2-\rho_0^2}\over\sqrt{\rho_1^2-\rho^2}},~~~~~~~~\rho_1 \equiv {1\over \sqrt{(1+B^2)}\,\omega}\eqno{(2.4)}$$
in which $\rho_0$ is an integration constant.

For a convenience, we plot a solution of (2.4) in figure 1 which represents a rotating D-string with 4 spikes.  The spiky D-string is rotating with angular velocity $\omega$ as could be read from (2.2).
\\
\\
\scalebox{1}{\hspace{5cm}\includegraphics{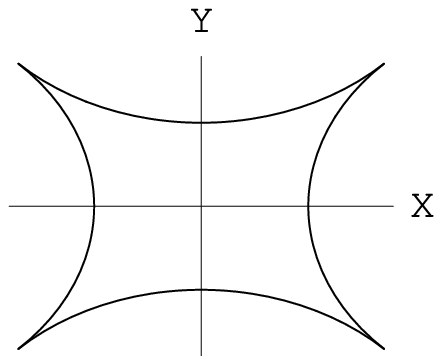}}
\\
\\
{\hspace{3cm} {\it Figure 1. A closed spiky D-string which has 4 spikes.  The D-string coordinate $\rho= \sqrt{x^2+y^2}$ is confined in the region $\rho_0 \le \rho \le\rho_1$.  It becomes spike at $\rho=\rho_1$ and has valleys at $\rho=\rho_0$ as $\rho' =0$ at this point. }
\\
\\
Using the above solution we can calculate the energy and angular momentum
$$E = P_t = {\sqrt\lambda\over 2\pi}\int d\sigma {\rho'^2+\rho^2\over \sqrt{\left(1-(1+B^2)\omega^2\rho^2\right)\rho'^2+\rho^2}} 
={\sqrt\lambda\over 4\pi} {\rho_1^2-\rho_0^2\over\rho_1^2}.\hspace{2cm}\eqno{(2.5)}$$
$$J = P_\theta = {\sqrt\lambda\over 2\pi}\int d\sigma {(1+B^2)\omega\rho^2\rho^2\over \sqrt{\left(1-(1+B^2)\omega^2\rho^2\right)\rho'^2+\rho^2}} 
={\sqrt\lambda\sqrt{1+B^2}\over 8\pi} \left({\rho_1^2-\rho_0^2}\right).\eqno{(2.6)}$$
The solution (2.4) tells us that the D-string coordinate $\rho$ is confined in the region $\rho_0 \le \rho \le\rho_1$.  It also shows the fact that $\rho' \rightarrow \infty$ if $\rho \rightarrow \rho_1$ and $\rho' \rightarrow 0$ if $\rho \rightarrow \rho_0$. Thus the D-string has  spikes at $\rho=\rho_1$ and has valleys at $\rho=\rho_0$ as $\rho' =0$ at this point.  Thus the geometric angle difference between the spike and valley is
$$\Delta \theta = \int d\sigma = \int d \rho\left({d \rho \over d\sigma}\right)^{-1}
=\int d \rho {\rho_0\over\rho_0\rho}{\sqrt{\rho_1^2-\rho^2}\over\sqrt{\rho^2-\rho_0^2}} = {\pi\over2}{\rho_1-\rho_0\over \rho_1}.\eqno{(2.7)}$$
In considering a close D-string which has $n$ spikes then $\Delta\theta = {\pi\over n}$. This implies that 
$$\rho_0 = \rho_1 \left(  1- {2\over n} \right).\eqno{(2.8)}$$
Substituting this relation into (2.5) and (2.6) the energy and angular momentum for a close D-string with  n spikes  become
$$E_n = {\sqrt \lambda\over 2\pi}\,2\rho_1 \left(1-{1\over n}\right)\hspace{1.5cm}.\eqno{(2.9)}$$
$$J_n = {\sqrt \lambda\over 2\pi}\,\rho_1^2\sqrt{1+ B^2} \left(1-{1\over n}\right).\eqno{(2.10)}$$
Above relations imply that 
$$E_n   = 2  \sqrt{{n-1\over n}\,J_n(B)},~~~~~J_n(B)\equiv {J\over \sqrt{1+B^2}}.\eqno{(2.11)}$$

Let us make two comments for the above analyses.
\\
1. The velocity at which the spikes move is described by
$$v= \rho_1{d\theta\over dt} = \rho_1{d\theta\over d\tau}{d\tau\over dt} = {1\over \sqrt{1+B^2}\,\omega} \omega = {1\over \sqrt{1+B^2}} < 1.\eqno{(2.12)}$$
Thus, in the case of B=0 the spike will move with the speed of light, as that found in [23].  However, the introducing external flux field will render it to slow down. 
\\
2.  In the case of  B=0 and n=2 the relation (2.11) recovers the standard Regge trajectory $E = \sqrt {2J}$, as that found in [23].  However, the introducing flux field will modify the relation in which there will appear an overall factor (which is the function of  $B$) in the angular momentum, i.e. $J\rightarrow J(B) ={1\over \sqrt{1+B^2}}\,J$.  We will see  that this property also shows in the following sections. 

%%%%%%%%%%%%%%%%%%%%%
\section {Giant Magnons under NS-NS Field}
The relevant part of the $AdS_3\times S^1$ metric and NS-NS B field  we used are described as [17]
$$ds^2 = -\cosh^2\rho dt^2 + d\rho^2+\sinh^2\rho d\chi^2 + d\phi^2,~~~~~{\bf B}= B\sinh^2\rho d\chi\wedge dt. \eqno{(3.1)}$$
To consider a rotating D-string we adopt the world-sheet coordinate [12] such that 
$$t = \tau, ~~~~~~\phi = t + \varphi(\sigma),~~~~~\chi= \omega\tau-\omega \psi(\sigma), ~~~~~\rho=\rho(\sigma).\eqno{(3.2)}$$
The Nambu-Goto Lagrangian of the rotating D-string is described by 
$${\cal L} = {\sqrt\lambda\over 2\pi}\int d\sigma \,\sqrt {\cal D},\eqno{(3.3)}$$
where
$${\cal D} = \cosh^2\rho\varphi'^2 +\omega^2\sinh^2\rho \cosh^2\rho \psi'^2 +(1-\omega^2)\sinh^2\rho \rho'^2 - \omega^2\sinh^2\rho(\varphi'+\psi)^2 - B^2\omega^2\sinh^4\rho \psi'^2.\eqno{(3.4)}$$
The associated energy and angular momenta of the rotating D-string are
$$E = P_t = {\sqrt\lambda\over 2\pi}\int {d\sigma \over {\sqrt {\cal D}}}
\left[\cosh^2\rho(\rho'^2+\omega^2\sinh^2\rho\psi'^2+\varphi'^2)-B^2\omega^2\sinh^4\rho\psi'^2\right],
\eqno{(3.5)}$$
$$S = P_\chi = {\sqrt\lambda\over 2\pi}\int {d\sigma \over {\sqrt {\cal D}}}
\left[\omega\sinh^2\rho(\rho'^2+\varphi'^2+\varphi'\psi')\right],\hspace{4.3cm}\eqno{(3.6)}$$
$$J = P_\phi = {\sqrt\lambda\over 2\pi}\int {d\sigma \over {\sqrt {\cal D}}}
\left[\rho'^2+\omega^2\sinh^2\rho(\psi'^2+\varphi'\psi')\right].\hspace{4.3cm}\eqno{(3.7)}$$
The Lagrangian implies the following relation
$$\psi' = {1\over (1-B^2)\sinh^2\rho}\, \varphi'\eqno{(3.8)}$$
Substituting this relation into (3.4) we have a simple form 
$${\cal D} = (1-\omega^2)\left(\cosh^2\rho \varphi'^2+\sinh^2\rho \rho'^2\right) - {\omega^2 B^2\over 1-B^2} \varphi'^2.\eqno{(3.9)}$$
The equation of motion of the D-string following from the Lagrangian then becomes
$${dr\over d\varphi} = {1\over\sqrt{r_1^2-r_0^2}}\sqrt{r^2-r_0^2}\sqrt{r^2-r_1^2}, ~~~~~r \equiv \cosh\rho,\eqno{(3.10)}$$
in which 
$$r_0^2 \equiv {\omega^2\over 1-\omega^2}{B^2\over 1-B^2},~~~~~~r^2_1\equiv r_0^2 + c^2,\eqno{(3.11)}$$
and $c$ is an integration constant. 

 The equation (3.10) may be regarded as a zero-energy particle of mass $M=2$ which is moving under a potential
$$V(r)= -{(r^2-r_0^2)(r^2-r_1^2)\over r_1^2-r_0^2}, \eqno{(3.12)}$$
in which the angle $\varphi$ is regarded as the ``time" variable.  The potential is plotted in figure 2.  We see that there are three typical D-string solutions:  1. If $r_1<1$, then  the rotating D-string extends over the region $1 \le r < \infty$.  2. If $r_1 \ge1$, then  the rotating D-string extends over the region $r_1 \le r < \infty$. 
3. If $r_0 > 1$, then  there is a rotating D-string extends over the finite region $ 1 \le r < r_0$. For a convenience, after solving (3.10) we plot the first two profiles of the D-string solution $r(\varphi)$ in figure 3.
\\
\\
\scalebox{1}{\hspace{5cm}\includegraphics{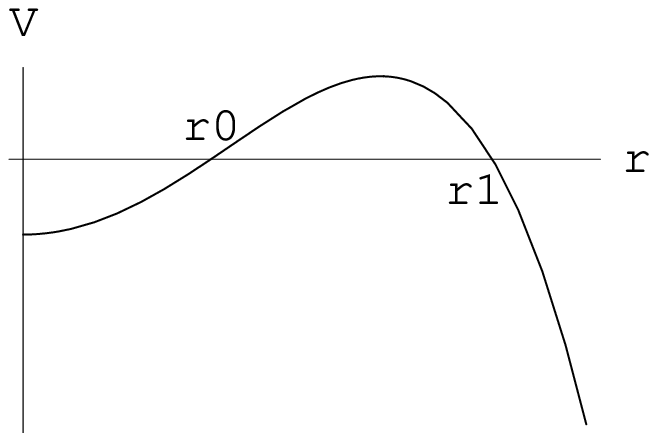}}
\\
\\
{\hspace{3cm} {\it Figure 2. The potential V(r) of (3.12) under which a particle is moving on. Only the case of $r1 <1$ could there is a giant magnon, which is a spiky D-string extends over the region $1 \le r < \infty$.}
\\

Using the relation (3.10) the energy and angular momentum of the rotating D-string could be simplified to be the simple forms
$$E-J={\sqrt \lambda\over \pi}{1\over \sqrt{1-\omega^2}}\int_{r_i}^{r_f} dr {\sqrt{r^2-r_0^2}\over\sqrt{r^2-r_1^2}}\hspace{8.5cm}$$
$$\hspace{2cm}+{\sqrt \lambda\over \pi}{B^2\over 1-B^2}{1\over \sqrt{1-\omega^2}}\left({1\over 1-\omega^2}-{\omega^2\over1-B^2}\right)\int_{r_i}^{r_f} dr {{r_1^2-r_0^2}\over\sqrt{r^2-r_1^2}(r^2-r_0^2)^{3/2}},\eqno{(3.13)}$$
$$S={\sqrt \lambda\over \pi}{\omega\over \sqrt{1-\omega^2}}\int_{r_i}^{r_f} dr {\sqrt{r^2-r_0^2}\over\sqrt{r^2-r_1^2}}\hspace{9cm}$$
$$+{\sqrt \lambda\over \pi}{B^2\over 1-B^2}{\omega\over \sqrt{1-\omega^2}}{1\over 1-\omega^2}\int_{r_i}^{r_f} dr {{r_1^2-r_0^2}\over\sqrt{r^2-r_1^2}(r^2-r_0^2)^{3/2}}.\eqno{(3.14)}$$
Note that $1 \le r \equiv \cosh\rho < \infty$.  Thus to regularize the integrations we have to introduce a cutoff  $\Lambda= r_f$  if the D-string extends to the infinity .
\\

  Let us first examine the situation of $B=0$, which is studied in [12].  Now
$r_0=0$ and we have the following two cases which have exactly analytic results. 
\\
\\
1. If $r_1<1$, then  the rotating D-string extends over the region $1 \le r < \infty$.  A possible D-string solution of (3.10) which has a spike at $r=1$ is shown in left-hand figure of figure 3.  In this case, we can use the (3.13) and (3.14) to calculate the energy and momentum.  The results are 
$$ E-J = {\sqrt\lambda\over \pi}{1\over \sqrt{1-\omega^2}}\left(\Lambda-\sqrt{1-r_1^2}\right),\hspace{8cm}\eqno{(3.15)}$$
$$~~~ S = {\sqrt\lambda\over \pi}{\omega\over \sqrt{1-\omega^2}}\left(\Lambda-\sqrt{1-r_1^2}\right)+{\sqrt\lambda\over \pi}{\omega\over (1-\omega^2)^{3/2}}\left({\sqrt{1-r_1^2}}+{1\over 2r_1}\left({\pi\over2}-\cos r_1\right)\right). \eqno{(3.16)}$$
\\
Then, using the relation $r_1= \cos\varphi_0$ (see (3.19)) and after subtracting the divergent terms the standard dispersion relation, $(E-J) =\sqrt{S^2 +{\lambda\over \pi^2}\sin^2{p\over2}}$\,, could be found.  This is the result first obtained in [12].
\\
\\
2. If $r_1>1$, then  the rotating D-string extends over the region $r_1 \le r < \infty$.  A possible D-string solution of (3.10) which is shown in left-hand figure of figure 3 has no spike and becomes smooth at $r=r_1$ because ${dr\over d \varphi} =0$ at this point.  In this case we can use the (3.13) and (3.14), while perform the integration from $r = r_1$ to $r= \infty$, to calculate the energy and momentum.   The results are 
$$ E-J = {\sqrt\lambda\over \pi}{\Lambda\over \sqrt{1-\omega^2}},~\hspace{4.5cm}\eqno{(3.17)}$$
$$~~~ S = {\sqrt\lambda\over \pi}{\omega\over \sqrt{1-\omega^2}}\,\Lambda+{\sqrt\lambda\over \pi}{\omega\over (1-\omega^2)^{3/2}}\,{\pi\over 4r_1}. \eqno{(3.18)}$$
\\
After subtracting the divergent terms the result, however, dose not show the dispersion relation of a magnon.   It thus seems that a spiky profile of a rotating D-string is crucial to map it to a magnon.  The details, however, remains to be investigated
\\
\\
\scalebox{1}{\hspace{5cm}\includegraphics{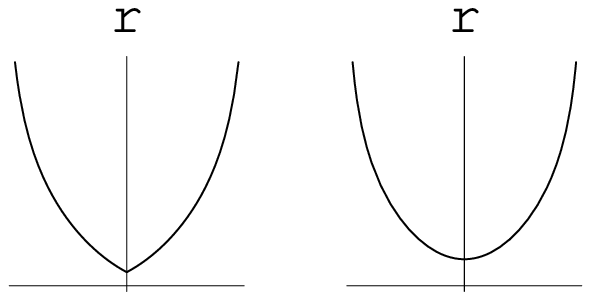}}
\\
\\
{\hspace{3cm} {\it Figure 3. Two D-string configurations $r(\varphi)$.  The left-hand figure is a D-string configuration extended over the region $1 \le r < \infty$, which has a spike and is dual to the magnon.  The right-hand figure is a D-string configuration extended over the region $1< r_1 \le r < \infty$, which has no spike and does not dual to the magnon. }
\\

 Let us now turn to the case of $B\ne 0$ and consider the D-string extended over the region $1 \le r < \infty$ to analyze the possible giant magnon solution. 

To proceed, let us calculate the geometrical angle of the giant magnon by using the equation (3.10).   However, an analytic relation could not be obtained without further approximation.  Therefore we will consider the case of small $B$ (Note that $r_0^2 \sim B^2$).   Thus 
$$\varphi_0 =\int d\varphi =\int_1^{\Lambda} dr\left({dr\over d \varphi}\right)^{-1}=\int_1^{\Lambda} d r {\sqrt{r_1^2 -r_0^2}\over \sqrt{r^2 -r_0^2}\sqrt{r^2 -r_1^2}}\approx \int_1^{\Lambda} dr {r_1\over r\,\sqrt{r^2 -r_1^2}}\left(1-{r_0^2\over 2r_1^2}+{r_0^2\over 2 r^2}\right)$$
$$=\left(1-{r_0^2\over 2r_1^2}\right) \left[\cos^{-1}(r_1) -{\pi\over2}\right]+ {r_0^2\over 2}\left[{\sqrt{1-r_1^2}\over2 r_1}+{1\over2 r^2_1}\left(\cos^{-1}(r_1) -{\pi\over2}\right)\right].\eqno{(3.19)}$$
For the small value of $r_0$ above equation has an approximation solution
$$r_1 \approx \left(1+ {r_0^2\over4}\left({\cos^2\varphi_0\over\sin^2\varphi_0}+{\cos\varphi_0\over\sin^2\varphi_0}\right)\right)\sin\varphi_0.\eqno{(3.20)}$$
In a similar way, in the case of small $B$ we can perform the integrations in (3.13) and (3.14) to find the analytic forms of the energy and angular momentum.  After substituting the relation (3.20) into the integrated forms we find that
$$(E-J)^2- S^2 = {\lambda\over \pi^2}\left[(1+B^2)\sin^2{p\over2}-B^2\sin{p\over2} - {B^2\over4} \left({\omega^2\over1-\omega^2}\sin^2{p\over2}\right)\cos^2{p\over2}\right]$$
$$~\approx {\lambda\over \pi^2}\left[(1+B^2)\sin^2{p\over2}-B^2\sin{p\over2} \right]- {B^2\over4}S^2\cos^2{p\over2} ,\eqno{(3.21)}$$
in which we have used the relation $S^2 = {\lambda\over \pi^2}{\omega^2\over1-\omega^2}\sin^2\varphi_0 +O(B^2)$.  Above relation implies the following dispersion relation
$$E-J \approx \sqrt{S(B)^2 + {\lambda\over \pi^2}\left[(1+B^2)\sin^2{p\over2}-B^2\sin{p\over2}\right]},\eqno{(3.22)}$$
where
$$S(B) \equiv\sqrt{\left(1- {B^2\over4}\cos^2{p\over2}\right)}\,\,S, \eqno{(3.23)}$$
in which the relation of the geometrical angle  $\varphi_0$ and the magnon momentum ${p}$ has been used. 
\\

 Let us make the following comments for the above result.
\\
1.  During the integrating we have introduced a cutoff  $\Lambda$ to regularize the integrations and adopt the prescription of [12] to add the counter terms to cancel the infinity to obtain a finite result.  This prescription has been found to be consistent and could reproduce a correct result, as detailed in [12].  Note that the infinity is coming from the infinite boundary of the AdS spacetime. 
\\
2. To obtain the final analytic result (3.22) we have considered the   approximation of  $B^2, \varphi_0 \ll  1$ and have used the replacement $B^2 \varphi_0 \rightarrow B^2 \sin \varphi_0$ while not the replacement $B^2 \varphi_0 \rightarrow \sin (B^2 \varphi_0)$.  This is because that the physical quantities S, J and E shall be a periodic function of $\varphi_0$.  Thus, the appearance of the phase $\sin (B^2 \varphi_0)$ will be unphysical unless it is coming from the periodic functional form of  $B^2\sin \varphi_0$. The derivation of a dispersion relation without these assumptions is an interesting work to be found.
\\
3. The dispersion relation of the giant magnon on the $\beta$ deformed $AdS_5 \times S^5$ found in [16,14] is $E-J =\sqrt{S^2 +{\lambda\over \pi^2}\sin^2\left({p\over2}-{\beta \pi}\right)}$ in which $\beta$ is a parameter of deformation.  Our result is more complex.  Especially, we have seen that in our model there is an overall factor before the angular momentum, i.e. $S \rightarrow S(B) \equiv\sqrt{\left(1- {B^2\over4}\cos^2{p\over2}\right)}\,S $.  The property had also shown in the comment 2 of section 2.  
\\
4. The result (3.23) shows that there is a pre-factor before the angular momentum $S$ but it does not be shown in the angular momentum $J$.  The asymmetric property may be traced to the fact that the external $B$ field is the tensor of  ${\bf B} \sim d\chi\wedge dt$ and $S \equiv P_{\chi}$ while  $J \equiv P_{\phi}$.  Note that the NS-NS B fields appear on $\beta$ deformed $AdS_5 \times S^5$ are symmetric in the angels of $\chi$ and $\phi$, thus it does not produce such an asymmetric property [16,14].
\\
5. Let us examine the case of $r_0 >1$.   Form figure 2 we see that there may exit a finite-size D-string which extended over the region $1 \le r <r_0$.  The solution has a spike and is expected to be able to map to a magnon.  However, from (3.13) and (3.14) we see that there is the factor $(r_0^2-r^2)^{-3/2}$ which will produce the divergence during integrating the variable  $r$ from $r=1$ to $r=r_0$ .   This divergence is not coming from the infinite boundary and could not be canceled by introducing a counter term.   Thus  the solution is unphysical and shall be neglected. 

%%%%%%%%%%%%%%%%%%%%%
\section {Giant Magnons under Melvin Field}
We first investigate the magnetic Melvin-field  effect on the giant magnon and then the electric Melvin-field  effect on the giant magnon. We follow the method in [11] in which the Polyakov action is adopted.
\subsection{Magnetic Melvin Field}
The relevant part of the metric and magnetic field we used are described by [21]
$$ds^2 = -\sqrt {1+ B^2\sin^2\theta}dt^2 + \sqrt {1+ B^2\sin^2\theta}d\theta^2+{\sin^2\theta\over\sqrt {1+ B^2\sin^2\theta}}d\phi^2 + \sqrt {1+ B^2\sin^2\theta} \cos^2\theta d\phi_1^2,$$
$$\hspace{3cm}A_\phi= {B\sin^2\theta\over 2 (1+ B^2\sin^2\theta)}. \eqno{(4.1)}$$
Due to the complex of the metric form we will in this section investigate the corresponding Polyakov Lagrangian of the rotating string which is described by 
$${\cal L} = - {\sqrt\lambda\over 4\pi}\int d\sigma \left(\sqrt{1+B^2(1-z^2)}+{\sqrt{1+B^2(1-z^2)}\over1-z^2}(-\dot z^2+z'^2)+ {(1-z^2)(-\dot\phi^2+\phi'^2)\over\sqrt{1+B^2(1-z^2)}}\right.$$
$$\hspace{3cm}\left. \sqrt{1+B^2(1-z^2)}\, z^2 (-\dot\phi_1^2+\phi_1'^2) + B z (\dot\phi\,z' -\dot z\, \phi' )\right),\eqno{(4.2)}$$
where we have used the approximation  $A_\phi \approx {B\sin^2\theta}$ which is in the case of small $B$ field and have defined the variable $z\equiv \cos\theta$. The energy and angular momentum of the rotating string are
$$E = P_\tau={\sqrt\lambda\over 2\pi}\int d\sigma \sqrt{1+B^2(1-z^2)},\hspace{2.5cm}\eqno{(4.3)}$$
$$J = P_\phi = {\sqrt\lambda\over 2\pi}\int d\sigma \left({(1-z^2)\dot\phi\over\sqrt{1+B^2(1-z^2)}}+ {1\over2}Bzz'\right).\eqno{(4.4)}$$
$$J_1 = P_{\phi_1} = {\sqrt\lambda\over 2\pi}\int d\sigma \sqrt{1+B^2(1-z^2)}z^2\, \dot\phi_1.\hspace{1.7cm}\eqno{(4.5)}$$ 

The Virasora constrains associated with the Lagrangian (4.2) are 
$${1\over1-z^2}(\dot z^2+z'^2)+ {(1-z^2)(\dot\phi^2+\phi'^2)\over {1+B^2(1-z^2)}}+ \, z^2 (\dot\phi_1^2+\phi_1'^2)= 1,\eqno{(4.6)}$$
$${1\over1-z^2}\dot z z'+ {(1-z^2) \dot\phi \phi'\over {1+B^2(1-z^2)}} + z^2 \dot\phi_1 \phi_1'= 0.~~~~~~~~~~\eqno{(4.7)}$$
Now, following the method of [11] we consider a two-spin spiky string in the following world-sheet coordinate 
$$t = \tau, ~~~~~~\phi = \omega t +\varphi(\sigma- v\omega \tau),~~~~~~\phi_1 = \nu \tau -\nu v \omega \sigma,~~~~~z= z(\sigma- v\omega \tau).\eqno{(4.8)}$$
It can be checked that the above ansatz satisfies the equation of motion and the Virasora constrains become the following relations
$${d\varphi\over d\sigma} = {v\over 1-v^2\omega^2}{\omega^2-B^2\over 1-z^2}~(z^2-z_{max}^2),\hspace{7cm}\eqno{(4.9)}$$
$$\left({dz\over d\sigma}\right)^2 = \left({\sqrt{1-B^2 v^2}\over \sqrt{1+B^2(1-z^2)}}{\sqrt{\omega^2-B^2}\over 1- v^2\omega^2}~\sqrt{z_{max}^2-z^2}~\sqrt{z^2-z_{min}^2}\right)^2 -\nu^2z^2(1-z^2),\eqno{(4.10)}$$
where
$$z_{min}^2 \equiv 1-{1\over \omega^2-B^2},~~~~~z_{max}^2\equiv1-{v^2\over 1-B^2v^2}. \eqno{(4.11)}$$
To proceed we find that it is difficult to find the analytic result and we will hereafter consider the one-spin spiky string by letting $\nu=0$.  Equations (4.9) and (4.10) then imply the following useful relation 
$${dz\over d\varphi} = {1\over \sqrt{1+B^2(1-z^2)}}{(1-z^2)\sqrt{1-B^2 v^2}\over v\,\sqrt{\omega^2-B^2}}~{\sqrt{z_{max}^2-z^2}\over \sqrt{z^2-z_{min}^2}},\eqno{(4.12)}$$
Using the relation (4.12) the energy and momentum could be expressed as a simple relation 
$$E-\omega J = {\sqrt \lambda\over \pi} \int_{z_{min}}^{z_{max}}dz ~{1\over\sqrt{1-B^2v^2}}{z\over\sqrt{z_{max}^2-z^2}} - {\sqrt \lambda\over \pi}\omega\,B(z_{max}^2-z_{min}^2). \eqno{(4.13)}$$
For a convenience, after solving (4.12) we plot a profile of the string solution $z(\varphi)$ in figure 4.
\\
\\
\scalebox{1}{\hspace{5cm}\includegraphics{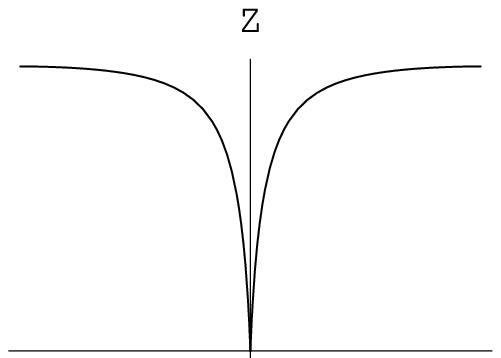}}
\\
\\
{\hspace{3cm} {\it Figure 4. A string configurations $z(\varphi)$.  The spiky string configuration extended over the region $ z_{min}\le z \le z_{max}$ is dual to the magnon. }
\\
\\

We now begin to calculate the geometrical angle of the giant magnon by using the equation (4.12).   However, an analytic relation could not be obtained without further approximation.  Therefore we will consider the case of small $B$.   Thus 
$$\varphi_0 = \int d\varphi=\int_{z_{min}}^{z_{max}}dz ~ \left({dz\over d\varphi}\right)^{-1}= \int_{z_{min}}^{z_{max}}dz \,v\,{\sqrt{1+B^2(1-z^2)}\sqrt{\omega^2-B^2}\sqrt{z^2-z_{min}^2}\over (1-z^2)\sqrt{1-B^2 v^2}\sqrt{z_{max}^2-z^2}}$$
$$\approx \int_{z_{min}}^{z_{max}}dz \left[{vz\over(1-z^2)\sqrt{z_{max}^2-z^2}}+{vB^2\over2}{z\over \sqrt{z_{max}^2-z^2}}\right]{1\over \sqrt{1-B^2 v^2}}\hspace{2cm}$$
$$\approx \cos^{-1}\left({v\over\sqrt{1-B^2v^2}}\right) + {vB^2\over2}.\hspace{7cm}\eqno{(4.14)}$$
Note that to obtain the above relation we have considered the limiting 
$$\omega^2\rightarrow 1+B^2.\eqno{(4.15)}$$
In this case $z_{min} \rightarrow 0$ and from (4.3) and (4.10) we see that 
$$E ={\sqrt\lambda\over 2\pi}\int dz \,\left({dz\over d\sigma}\right)^{-1} \sqrt{1+B^2(1-z^2)},\hspace{4.5cm}$$
$$ = {\sqrt\lambda\over \pi}(1- v^2\omega^2)\int_0^{z_{max}} {dz\over z} \,\,{1+B^2(1-z^2)\over\sqrt{1-B^2 v^2}}~{1\over \sqrt{z_{max}^2-z^2}}\rightarrow \infty,\eqno{(4.16)}$$
In the same way  $J \rightarrow \infty$.  Thus, in the limit of (4.15) we recover the giant magnon found by Maldacena and Hofman [7]. 

  For a small value of $B$ field the equation (4.14) has an approximation solution
$$v \approx (1-B^2 \cos^2 \varphi_0)\cos\left(\varphi_0 -{B^2\over2}\cos\varphi_0\right).\eqno{(4.17)}$$
Using this relation the equation (4.13) becomes
$$E - J(B) \approx {\sqrt \lambda\over \pi}\left[\sin{p\over 2}+{B^2\over2}\left(\cos^2{p\over 2}\sin{p\over 2}-\cos^2{p\over 2}\right)- {B\over2} \sin^2{p\over 2}\right],,\eqno{(4.18)}$$
where 
$$J(B)\equiv \sqrt{1+B^2}\, J,\eqno{(4.19)}$$
in which we have replaced the geometrical angle  $ 2\varphi_0$ by the magnon momentum ${p}$.  Thus we have found the correction terms of the giant magnon under a magnetic Melvin field.

  Let us make the following comments for the above result.
\\
1.  We see that there is an overall factor before the angular momentum, i.e. $J\rightarrow J(B)\equiv  \sqrt{1+B^2}\, J$.  The property had shown in the sections II and III.
\\
2.   In the AdS/CFT [1-3] or string/spin chain [24] correspondence one maps the giant magnons to the magnon excitations of low lying spin chain states.   However, under the magnetic field background the corresponding spin has been deformed and more interactions are added, as shown in our previous paper [25] and in [26]. Thus the corresponding dispersion relation will show this properties in the correction terms.  Searching a corresponding spin chain system and making the precise map between the spin deformations and the  correction terms in the dispersion relation is an interesting problem and worthy to study. 
\\
3.  In principle we can also study the giant magnon in the SL(2) sector for the string moving on the magnetic-field deformed $AdS_3\times S^1$.  The background constructed by us in [22] is just (4.1) while with the replacements $\theta \rightarrow \rho$ and $\sin^2\theta \rightarrow \sinh^2\rho$.  However, in this case there will appear the factor $B^2 \sinh^2\rho$ and because $\sinh^2\rho\rightarrow \infty$ as $\rho \rightarrow \infty$ we are unable to analyze the system of small $B$ field.

%%%%%%%%%%%%%%%%%%%%%%%%
\subsection{Electric Melvin Field}
The relevant part of the metric and electric field we used are described by [21]
$$ds^2 = -{dt^2\over \sqrt {1 - {\cal E}^2}} + \sqrt{1 - {\cal E}^2}\left(d\theta^2+\sin^2\theta d\phi^2 \right),\hspace{2cm} A_t = {{\cal E}\over\sqrt {1 - {\cal E}^2}}. \eqno{(4.20)}$$
The corresponding Polyakov Lagrangian of the rotating string is described by 
$${\cal L} = - {\sqrt\lambda\over 4\pi}\int d\sigma \left({1\over \sqrt {1 - {\cal E}^2}}+\sqrt {1 - {\cal E}^2}\,\,{-\dot z^2+z'^2\over1-z^2}+\sqrt {1 - {\cal E}^2} \, (-\dot\phi^2+\phi'^2)\right),\eqno{(4.21)}$$
where we have defined the variable $z\equiv\cos\theta$. The energy and angular momentum of the rotating string are
$$E = P_\tau={\sqrt\lambda\over 2\pi}\int d\sigma {1\over \sqrt {1 - {\cal E}^2}},\hspace{1.8cm}\eqno{(4.22)}$$
$$J = P_\phi = {\sqrt\lambda\over 2\pi}\int d\sigma \,\sqrt {1 - {\cal E}^2}\, (1-z^2)\dot\phi.\eqno{(4.23)}$$

The Virasora constrains associated with the Lagrangian (4.21) are 
$$\sqrt {1 - {\cal E}^2} \,{-\dot z^2+z'^2\over1-z^2}+\sqrt {1 - {\cal E}^2} \, (-\dot\phi^2+\phi'^2)={1\over \sqrt {1 - {\cal E}^2}},\eqno{(4.24)}$$
$${\dot z z'\over1-z^2}+ (1-z^2) \dot\phi \phi'= 0.\eqno{(4.25)}$$
Now, as before, we consider a spiky string moving in the following world-sheet coordinate 
$$t = \tau, ~~~~~~\phi = \omega t +\varphi(\sigma- v\omega \tau),~~~~~z= z(\sigma- v\omega \tau).\eqno{(4.26)}$$
It can be checked that the above ansatz satisfies the equation of motion and the Virasora constrains become the following relations
$${d\varphi\over d\sigma} = {v\omega^2\over 1-v^2\omega^2}{z^2-z_{max}^2\over 1-z^2},\hspace{2.6cm}\eqno{(4.27)}$$
$$\left({dz\over d\sigma}\right)^2 = {\omega^2\over \left(1- v^2\omega^2\right)^2}~\left(z_{max}^2-z^2\right)\,\left(z^2-z_{min}^2\right),\eqno{(4.28)}$$
where
$$z_{min}^2 \equiv 1-{1\over \left(1-{\cal E}^2\right)\omega^2},~~~~~z_{max}^2\equiv1-{v^2\over 1-{\cal E}^2}. \eqno{(4.29)}$$
Equations (4.28) and (4.29) imply the following useful relation 
$${dz\over d\varphi} = {(1-z^2)\over v\omega}~{\sqrt{z_{max}^2-z^2}\over \sqrt{z^2-z_{min}^2}},\eqno{(4.30)}$$
After solving (4.30) we can plot a profile of the string solution $z(\varphi)$.  The result is like that in figure 4.

Now, using the relation (4.28) the energy (4.22) could be expressed as a simple relation 
$$E ={\sqrt\lambda\over 2\pi}\int dz \,\left({dz\over d\sigma}\right)^{-1} {1\over \sqrt {1 - {\cal E}^2}},\hspace{4.5cm}$$
$$ = {\sqrt\lambda\over \pi}{1- v^2\omega^2\over \omega\,\sqrt {1 - {\cal E}^2}}\int_{z_{min}}^{z_{max}} \,dz ~{1\over \sqrt{z^2- z_{min}^2}}~{1\over \sqrt{z_{max}^2-z^2}}.\eqno{(4.31)}$$
In the same way  the angular momentum (4.23) becomes
$$J  =  {\sqrt\lambda\over \pi}{1\over \sqrt {1 - {\cal E}^2}}\int_{z_{min}}^{z_{max}} \,dz ~{\sqrt{z_{max}^2-z^2}\over \sqrt{z^2- z_{min}^2}}.\eqno{(4.32)}$$
Thus, in the limit of 
$$ \omega = {1\over \sqrt{1-{\cal E}^2}},\eqno{(4.33)}$$
we have the property 
$$z_{min}=0,~~~~~~~~\Rightarrow ~~~~~~E,\, J \rightarrow \infty,\eqno{(4.34)}$$
and we recover the giant magnon found by Maldacena and Hofman [7].  It is also easy to see that in the limit (4.33) we have a simple relation
$$E-\omega J = E- {J\over\sqrt{1-{\cal E}^2}}= {\sqrt \lambda\over \pi}\int_0^{z_{max}}dz ~{z\over\sqrt{z_{max}^2-z^2}}=  {\sqrt \lambda\over \pi}\,z_{max}. \eqno{(4.35)}$$

To proceed we have to calculate the geometrical angle of the giant magnon.  In the limit of (4.33) the equation (4.30) implies
$$\varphi_0 = \int d\varphi=\int_{z_{min}}^{z_{max}}dz ~ \left({dz\over d\varphi}\right)^{-1}= {v\over\sqrt{1-{\cal E}^2}}\,\int_0^{z_{max}}dz \,{z\over (1-z^2)\sqrt{z_{max}^2-z^2}}= \cos^{-1}\left({v\over \sqrt{1-{\cal E}^2}}\right). \eqno{(4.36)}$$
Thus 
$$v = \sqrt{1-{\cal E}^2}\, \cos\varphi_0.\eqno{(4.37)}$$
Using this relation the equation (4.35) becomes
$$E- J({\cal E})= {\sqrt \lambda\over \pi}\,\sin{p\over 2}, ~~~~~~~~~~J({\cal E})\equiv {J\over\sqrt{1-{\cal E}^2}}, \eqno{(4.38)}$$
in which we have replaced the geometrical angle  $ 2\varphi_0$ by the magnon momentum ${p}$.  We thus have found an interesting property that the effect of an electric Melvin field on the dispersion relation of the giant magnon is merely to add a prefactor before the angular momentum, i.e. $J \rightarrow J({\cal E})\equiv{J\over\sqrt{1-{\cal E}^2}}$.   Note that, while the relations (3.22) and (4.18) are the results under a small $B$ field the dispersion relation (4.38) is obtained without any approximation.
\\

Finally, let us make the following comments to conclude this paper.
\\
1. The property we found in this paper is that the corrections of the dispersion relation of a giant magnon under the external NS-NS or Melvin field are to appear a prefactor before the angular momentum, in addition to some corrections in the sine function.
\\
2. For the giant magnons in the $\beta$ deformed  $Ads_5\times S^5$ it is known that the new dispersion relation merely correct the phase factor, i.e. $\sin\left({p\over2}\right) \rightarrow \sin\left({p\over2}-{\beta \pi}\right)$ [16,14].  However, as $\sin\left({p\over2}-{\beta \pi}\right) = \sin\left({p\over2}\right)  \cos\left({\beta \pi}\right)  - \cos\left({p\over2}\right)  \sin\left({\beta \pi}\right) $ it may be regarded as a specific form of the correction terms in the sine function.
\\
3. It is also interesting to consider a spiky string moving in the background 
$$ds^2 = -dt^2 + d\theta^2 + \sin^2\theta d\phi^2 ,\hspace{2cm} B_{\theta\phi} = B\,\sin\theta,  \eqno{(4.39)}$$
which is the spacetime used by Bachas et al  to study the flux stabilization of D-branes [27].  Then, following the method of this section we can easily show that the dispersion relation of the corresponding giant magnon becomes
$$E- J = {\sqrt \lambda\over \pi}\,(1+ B)\,\sin{p\over 2}.\eqno{(4.40)}$$
In this case there appears a correction in the sine function, i.e. $\sin{p\over 2} \rightarrow (1+ B)\,\sin{p\over 2}$ and it does not show any prefactor before the angular momentum.  
%%%%%%%%%%%%%%%%%%%%%
\section{Discussions}
Since Maldacena and Hofman [7] identified the elementary magnon with a rotating spiky string configuration moving on an $R \times S^2$, which they called the `giant magnon.'   there are several literatures had investigated the properties of the corresponding string solutions [10-16].  In this paper we have  investigated the effects of the NS-NS and Melvin fields on the giant magnon. We first consider the two-spin spiky D-string moving on the $AdS_3\times S^1$ with NS-NS antisymmetric B field background.  We see that there are D-string solution with or without spike, which is extended to the infinite boundary of the de-Sitter spacetime.  After the regularization we see that only those with spikes could be mapped to the magnon.  For these solutions which extended to infinity we add the counter term to cancel the divergence in the energy and momentum and have obtained the dispersion relation of the corresponding giant magnon.  The result shows how the background field will produce the correction terms.   We also investigate the spiky fundamental string moving on the electric or magnetic Melvin field background.  As the Melvin fields couple differently to particles of different spins they naturally break supersymmetry.  Therefore the correspondence between the magnon and string will be in the content of less supersymmetry, which is interesting from  the point of view of phenomena. The dispersion relation of the corresponding Melvin-field deformed giant magnon is also obtained.   We discuss some properties of the correction terms and their relations to the spin chain with deformations.   
%%%%%%%%%%%%%%%%%%%%%%
\\
~
\\
~
\\
\begin{center}{\Large \bf  References}\end{center}
\begin{enumerate}
\item J.~M.~Maldacena, ``The large N limit of superconformal field theories and supergravity,'' Adv.\ Theor.\ Math.\ Phys.\  {\bf 2}, 231 (1998) [hep-th/9711200].
\item S.~S.~Gubser, I.~R.~Klebanov and A.~M.~Polyakov, ``Gauge theory correlators from non-critical string theory,'' Phys.\ Lett.\ B428 (1998) 105 [hep-th/9802109]; E.~Witten, ``Anti-de Sitter space and holography,'' Adv.\ Theor.\ Math.\ Phys.\   2 (1998) 253 [hep-th/9802150].
\item  O.~Aharony, S.~S.~Gubser, J.~M.~Maldacena, H.~Ooguri and Y.~Oz, ``Large N field theories, string theory and gravity,'' Phys.\ Rept. 323 (2000) 183 (2000) [hep-th/9905111]; E.~D'Hoker and D.~Z.~Freedman, ``Supersymmetric gauge theories and the AdS/CFT correspondence,''  [hep-th/0201253].
\item S.~Frolov and A.~A.~Tseytlin, ``Multi-spin string solutions in
$AdS_5 \times S^5$,'' Nucl.\ Phys.\ B668 (2003) 77 [hep-th/0304255]; ``Quantizing three-spin string solution in $AdS_5 \times S^5$,'' JHEP 0307 (2003) 016 [hep-th/0306130];  R.C.Rashkov and K.S.Viswanathan, ``Rotating Strings with B-field,'' [hep-th/0211197 ].
\item A.~A.~Tseytlin, ``Spinning strings and AdS/CFT duality,'' [hep-th/0311139].
\item J, Plefka, ``Spinning strings and integrable spin chains in the AdS/CFT correspondence,'' [hep-th/0507136].
\item  D. M. Hofman and J. M. Maldacena, ``Giant magnons," J.Phys. A39 (2006) 13095-13118[hep-th/0604135]. 
\item  K. Pohlmeyer, ``Integrable Hamiltonian Systems And Interactions Through Quadratic Constraints," Commun. Math. Phys. 46, 207 (1976);  A. Mikhailov, ``An action variable of the sine-Gordon model,"  [hep-th/0504035];  A. Mikhailov, ``Baecklund transformations, energy shift and the plane wave limit," 
[hep-th/0507261]; A. Mikhailov, ``A nonlocal Poisson bracket of the sine-Gordon model,"  [hep-th/0511069]. 
\item G. Arutyunov, S. Frolov and M. Staudacher, ``Bethe ansatz for quantum strings," JHEP 0410, 016 (2004) [hep-th/0406256]. 
\item N. Dorey, ``Magnon bound states and the AdS/CFT correspondence," J.Phys. A39 (2006) 13119-13128 [hep-th/0604175];  H. Y. Chen, N. Dorey and K. Okamura, ``Dyonic giant magnons," JHEP 0609 (2006) 024 [hep-th/0605155]. 
\item G. Arutyunov, S. Frolov, and M. Zamaklar, ``Finite-size Effects from Giant Magnons," [hep-th/0606126].
\item J.A. Minahan, A. Tirziu, A.A. Tseytlin, ``Infinite spin limit of semiclassical string states," JHEP 0608 (2006) 049 [hep-th/0606145].
\item M. Spradlin and  A. Volovich, ``Dressing the Giant Magnon," [hep-th/0607009]; M. Kruczenski, J. Russo, and A.A. Tseytlin, ``Spiky strings and giant magnons on S5," [hep-th/0607044].
\item N. P. Bobev and  R. C. Rashkov, ``Multispin Giant Magnons," Phys.Rev. D74 (2006) 046011 [hep-th/0607018] ;  S. Ryang, ``Three-Spin Giant Magnons in $AdS_5 \times S^5$," [hep-th/0610037].
\item  P. Bozhilov and  R. C. Rashkov, ``Magnon-Like Dispersion Relation from M-theory," [hep-th/0607116].
\item C. S. Chu, G. Georgiou, and V. V. Khoze, ``Magnons, Classical Strings and beta-Deformations," [hep-th/0606220].
\item C.~Bachas and M.~Petropoulos, ``Anti-de-Sitter D-branes'', JHEP  0102 (2001) 025 [hep-th/0012234].
\item P. M. Petropoulos, S. Ribault,  ``Some comments on Anti-de Sitter D-branes'' , JHEP 0107 (2001) 036 [hep-th/0105252];
A.~Giveon, D.~Kutasov and A.~Schwimmer, ``Comments on D-branes in AdS(3)'', Nucl.\ Phys.\ B {\bf 615}, 133 (2001) [hep-th/0106005]; 
Y. Hikida and Y. Sugawara, ``Boundary States of D-branes in $AdS_3$ Based on Discrete Series'', Prog. Theor.Phys. 107 (2002) 1245-1266 [ hep-th/0107189];
A.~Rajaraman and M.~Rozali, ``Boundary states for D-branes in AdS(3),'' Phys. Rev.\ D {\bf 66}, 026006 (2002) [hep-th/0108001];
P.~Lee, H.~Ooguri and J.~w.~Park, ``Boundary states for AdS(2) branes in AdS(3),'' Nucl.\ Phys.\ B {\bf 632}, 283 (2002)  [hep-th/0112188].
\item  C. Deliduman, ``$AdS_2$ D-Branes in Lorentzian $AdS_3$'',  Phys.Rev. D68 (2003) 066006 [hep-th/0211288]; J.~Kumar and A.~Rajaraman, ``Revisiting D-branes in $AdS_3 \times S^3$'', Phys.Rev. D70 (2004) 105002 [hep-th/0405024]; D. Israel, ``D-branes in Lorentzian AdS(3)'', [hep-th/0502159];  Wung-Hong Huang, ``Anti-de Sitter D-branes in Curved Backgrounds,'' JHEP 0507 (2005) 031 [hep-th/0504013]. 
\item  O.~Lunin and J.~Maldacena, ``Deforming field theories with U(1) x U(1) global symmetry and their gravity duals,'' JHEP  0505  (2005)  033  [hep-th/0502086]; S.A. Frolov, R. Roiban, A.A. Tseytlin, ``Gauge-string duality for superconformal deformations of N=4 Super Yang-Mills theory,'' JHEP 0507 (2005) 045 [hep-th/0503192 ]; S. Frolov, ``Lax Pair for Strings in Lunin-Maldacena Background,'' JHEP 0505 (2005) 069 [hep-th/0503201]; R.~G.~Leigh and M.~J.~Strassler, ``Exactly marginal operators and duality in four-dimensional N=1 supersymmetric gauge theory,'' Nucl.\ Phys.\ B447 (1995) 95 [hep-th/9503121].
\item  Wung-Hong Huang, ``Semiclassical Rotating Strings in  Electric and Magnetic Fields Deformed $AdS_5 \times S^5$ Spacetime'', Phys. Rev. D73 (2006) 026007 [hep-th/0512117]; Wung-Hong Huang, ``Electric/Magnetic Field Deformed Giant Gravitons in Melvin Geometry'', Phys.Lett. B635 (2006) 141 [hep-th/0602019 ]; Wung-Hong Huang, ``Spinning String and Giant Graviton in Electric/Magnetic Field Deformed $AdS_3 \times S^3 \times T^4$'', Phys.Rev. D73 (2006) 126010 [hep-th/0603198 ]. 
\item  Wung-Hong Huang, ``Multi-spin String Solutions in Magnetic-flux Deformed $AdS_n \times S^m$ Spacetime'', JHEP 0512 (2005) 013 [hep-th/0510136 ].
\item   M. Kruczenski,  ``Spiky strings and single trace operators in gauge theories,''   JHEP  0508 (2005) 014 [hep-th/0410226];  S. Ryang, ``Wound and Rotating Strings in $AdS_5 \times S^5$,'' JHEP 0508 (2005) 047 [hep-th/0503239 ].
\item M.~Kruczenski, ``Spin chains and string theory,'' Phys. Rev. Lett. 93 (2004) 161602. h[ep-th/0311203];   R.~Hernandez and E.~Lopez, ``The SU(3) spin chain sigma model and string theory,'' JHEP 0404 (2004) 052 [hep-th/0403139]; S. Bellucci, P. Y. Caesteill, J. F. Morales and C. Sochichi, ``SL(2) spin chain and spinning strings on $AdS_5 \times S^5$,'' Nucl.\ Phys.\ B707 (2005) 303 [hep-th/0409086]; S.~A.~Frolov, R.~Roiban and A.~A.~Tseytlin, ``Gauge-string duality for superconformal deformations of N = 4 super Yang-Mills theory,'' JHEP 0507 (2005) 045 [hep-/0503192]; H. Dimov, R.C. Rashkov, ``A note on spin chain/string duality,'' Int.J.Mod.Phys. A20 (2005) 4337 [hep-th/0403121 ].
\item  Wung-Hong Huang, ``Spin Chain with Magnetic Field and Spinning String in Magnetic Field Background'', Phys. Rev. D74 (2006) 027901
 [hep-th/0605242]. 
\item  Wen-Yu Wen, ``Spin chain from marginally deformed $AdS_3 \times S^3$,'' [hep-th/0610147 ].
\item  C. Bachas, M. Douglas, C. Schweigert, ``Flux Stabilization of D-branes,'' JHEP 0005 (2000) 048 [hep-th/0003037].

\end{enumerate}
\end{document}